\documentclass{vgtc}                          % final (conference style)
\usepackage{CJK}
\graphicspath{{figures/}{pictures/}{images/}{./}} % where to search for the images

\usepackage{times}                     % we use Times as the main font
\usepackage{tabu}                      % only used for the table example
\usepackage{booktabs}                  % only used for the table example
\usepackage{lipsum}                    % used to generate placeholder text
\usepackage{mwe}                       % used to generate placeholder figures

\usepackage{mathptmx}                  % use matching math font

\vgtcinsertpkg

\title{Vibe Coding for Visualization Implementation: An Empirical Study of Practices and Challenges}

\author{Zhengyu Sun\thanks{e-mail: zhengyu003@e.ntu.edu.sg}\\ %
        \scriptsize Nanyang Technological University %
\and Xiaolin Wen\thanks{e-mail: xiaolin004@e.ntu.edu.sg}\\ %
     \scriptsize Nanyang Technological University %
\and Fengjie Wang\thanks{e-mail: fengjie.wang@connect.ust.hk}\\ %
     \scriptsize The Hong Kong University of Science and Technology %
\and Can Liu\thanks{e-mail: can.liu@ntu.edu.sg}\\ %
     \scriptsize Nanyang Technological University %
\and Yi Lai\thanks{e-mail: laiy0038@e.ntu.edu.sg}\\ %
     \scriptsize Nanyang Technological University %
\and Christophe Hurter\thanks{e-mail: christophe.hurter@enac.fr}\\ %
     \scriptsize ENAC, Universit{\'e} de Toulouse %
\and Yong Wang\thanks{e-mail: yong-wang@ntu.edu.sg}\\ %
     \scriptsize Nanyang Technological University}

\abstract{
Data visualization is essential for data analysis and communication, yet creating expressive visualizations remains labor-intensive. Recent AI-driven ``vibe coding'' tools enable users to generate visualizations through natural language interaction, lowering the barrier to entry. However, visualization implementation requires precise alignment between user intent and visual representation, which may differ from general software development practices. We present an empirical study with 16 participants of varying expertise to examine how users employ vibe coding tools for visualization implementation. Participants completed two visualization tasks and a semi-structured interview. Our findings characterize the diverse practices users adopt across prompting, evaluation, and iteration, and surface the challenges they encounter throughout the process. 
} % end of abstract


\definecolor{RED}{rgb}{0.7,0.0,0.0}

\newcommand{\wy}[1]{\textcolor{brown}{[WY: #1]}}

\newcommand{\ff}[1]{\textcolor{RED}{[FF: #1]}}

\newcommand{\szy}[1]{\textcolor{teal}{[SZY: #1]}}

\begin{document}

\begin{CJK}{UTF8}{gbsn}



% \firstsection{Introduction}

\maketitle

\section{Introduction}
% V2.1

% \ff{(prompt for AI) if you notice any [refs], [ref] alike marks, they are the placeholders of the citations I want to cover, please locate proper and representative papers to replace the placeholders, attach their bib in your response, add at most 3 papers for each placeholder.}

% \ff{we not only target developers, also include designers who lack of coding expertise, and general people with the needs for (customized) visualization creation, so adjust your writing accordingly.}
Data visualization plays a central role in data analysis and decision-making~\cite{wen2025prettismart,wang2021dodrio}, enabling users to explore and communicate insights from data.
Traditionally, creating an expressive visualization is labor-intensive: practitioners either write D3~\cite{bostock2011d3} code line by line or rely on interactive tools (e.g., Tableau) to coordinate data transformation, visual encoding, and interaction.
Recently, tools such as Codex~\cite{openai_codex_2026} and Antigravity~\cite{google_antigravity_2025} have introduced a new implementation paradigm known as ``vibe coding''~\cite{sarkar2025vibe}, in which users create artifacts through dialogue with AI rather than by hand.
This paradigm has been rapidly adopted, opening the activity to a broader audience that spans experienced developers and general users without coding expertise.

% \ff{reframe the arguments here to distinguihs from previous work:
% 1. how visualization implementation tasks different from general software development tasks
% 2. focus of visualization community, more about harnessing the power of llm to support nl-based workflow to author visualiztaion [check, adjuts and enrich the argument], but lack empirical understanding of how users with different expertise make use of this new paradigm, specially for some customized visualization, which is often the needs of info designers, and [also list other scenarios]
% }
However, visualization implementation poses demands that differ from general software development.
In typical software tasks, a single intent often admits many acceptable implementations~\cite{peleg2018programming}, and developers tend to accept AI output when it ``gets close'' to the intended functionality and then refine from there~\cite{sarkar2025vibe}.
% \ff{since we already mention data transformation, visual encoding and interaction elesewhere, the statement should be refined accordingly to align with this, as now, it's focusing on visual encoding.}
Visualization implementation works differently: the user's intent is to convey a specific data insight, and fine-grained choices across data transformation, visual encoding, and interaction jointly determine whether that insight is communicated effectively~\cite{card1999readings}, often requiring careful iterative tuning.
Meanwhile, although the visualization community has actively explored LLMs for natural language-based authoring workflows~\cite{dibia2023lida,tian2024chartgpt,wang2025data}, we still lack an empirical understanding of how users with different expertise actually engage with vibe coding tools, particularly when producing customized visualizations commonly needed by information designers and domain analysts.
We hypothesize that, in the visualization context, users prompt and evaluate AI output in ways that diverge meaningfully from general software development, warranting dedicated empirical investigation.

To address this gap, we conducted an empirical study with 16 participants of varying expertise, guided by two research questions:
(RQ1) how do users with different expertise levels leverage vibe coding tools to create visualizations?
and (RQ2) what challenges do they encounter throughout the process of using vibe coding tools?
Each participant completed a demographic questionnaire, two 30-minute visualization implementation tasks, and a semi-structured interview.
% \ff{also highight our design consideration here, e.g., mimic the scenario when designers have an idea of the desired visualization, then make use of vibe coding to create their intended visualization.}
The tasks were designed to mimic a realistic scenario in which a user has a target visualization in mind and uses vibe coding to realize it, covering representative aspects of visualization implementation including data transformation, visual encoding, and interaction. 
Based on our analysis, we present a comprehensive understanding of how users integrate vibe coding into their workflows, highlighting the specific practices they adopt for iterative refinement in human-AI collaboration. Furthermore, we identify the key challenges encountered during this process.
Finally, drawing on these insights, we discuss the implications for designing future vibe coding tools that better support user-centric, collaborative visualization implementation.

\section{Related Work}
% V2.1

Visualization authoring tools have steadily lowered the technical barrier to creating expressive charts. Low-level programming libraries such as D3~\cite{bostock2011d3} offer fine-grained control over visual primitives, enabling highly customized designs at the cost of substantial coding effort.
% and a steep learning curve. 
To reduce this burden, declarative languages (e.g., ggplot2~\cite{wickham2011ggplot2}, Vega-Lite\cite{satyanarayan2016vega}, Altair\cite{vanderplas2018altair}) have been introduced, which describe visualizations as mappings from data columns to visual channels, abstracting away the linkage between data items and visual objects. 
Building on these grammars, interactive tools such as Lyra~\cite{satyanarayan2014lyra}, Data Illustrator~\cite{liu2018data}, Charticulator\cite{ren2018charticulator}, and Tableau~\cite{tableau} expose shelf-configuration interfaces for specifying visual encodings.
% while recommendation systems like Voyager[refs], Lux[refs], and Draco[refs] suggest charts from partial specifications. 
These higher-level abstractions accelerate authoring but constrain users to a predefined design space and typically assume tidy data~\cite{wickham2014tidy}, trading expressiveness for accessibility.

To further lower the entry barrier, AI-powered tools translate natural-language (NL) queries into visualizations.
Early systems relied on rule-based parsing (e.g., NL4DV~\cite{narechania2020nl4dv}, Articulate~\cite{sun2010articulate}) or machine learning-based translation (e.g., RGVisNet~\cite{song2022rgvisnet}), but supported only a limited chart vocabulary and demanded precise queries or low-level demonstration examples.
The code-generation capabilities of large language models (LLMs) have since motivated a new class of NL2VIS tools~\cite{dibia2023lida,maddigan2023chat2vis,tian2024chartgpt} that accept free-form NL input.
For instance, 
LIDA~\cite{dibia2023lida} summarizes a dataset and prompts an LLM to generate Python code for chart creation; 
Chat2Vis~\cite{maddigan2023chat2vis} embeds visualization-specific prompts to improve generation reliability; 
ChartGPT~\cite{tian2024chartgpt} decomposes the task into a fine-grained reasoning pipeline (e.g., column selection, filtering, chart type selection, visual encoding) via chain-of-thought prompting.
Beyond these, researchers have also explored interaction-enhanced instruction~\cite{shen2025prompting, shen2026interaction} and retrieval augmentation~\cite{IntelliCircos} techniques to better utilize the power of LLMs.
For example, DynaVis~\cite{vaithilingam2024dynavis} synthesizes dynamic, on-demand UI widgets in response to user editing intent of the generated visualization.
% while benchmarks such as VisEval [ref] and the systematic study by Vazquez et al. [ref] characterize recurring LLM failure modes across chart generation, library adaptation, and visual encoding.

% While these LLM-based NL2VIS tools share with vibe coding a reliance on natural-language prompting, they still scaffold users within structured pipelines and largely predetermined output formats.
The emergence of \emph{vibe coding} marks a more open-ended turn in this trajectory.
In early 2025, Karpathy introduced the term to describe a workflow in which users converse with an AI ``composer'' to develop software, often without inspecting the underlying code; the paradigm has since been popularized by tools such as Cursor, Windsurf, and Trae, which shift the user's focus from code manipulation to high-level intent.
% Ge~\cite{ge2025vibe} subsequently characterized vibe coding as a software development method that leverages LLMs.
Applied to visualization authoring, vibe coding loosens the constraints of prior NL2VIS pipelines: users can iteratively produce visualization code in any library or output format through multimodal, conversational interaction. %\wy{double-check whether this contrast with NL2VIS tools is accurate, since some recent tools also allow flexible code output.}

Despite these technical advances, empirical understanding of how people actually use LLMs, and vibe-coding workflows in particular, for visualization authoring remains limited. 
% \ff{please supplement with current empirical studies in this context (e.g., user studies on NL2VIS tools, expertise effects, or task-complexity effects) to better support this gap claim.} 
% \ff{summarize current related empirical study around vibe coding/llm for visualization authoring and point out the gap, based your decision on recent literature in visualization and hci domain.} 
Existing investigations of vibe coding have largely centered on general software development, leaving open how it manifests in visualization tasks, how users with different levels of expertise leverage it, and how their interaction patterns evolve as task complexity grows. 
Our work addresses this gap.

\section{User Study}

To address the research questions introduced above, we conducted an empirical user study.
Each participant received 15 SGD per hour as compensation.
Each session followed a three-phase protocol: a briefing, the visualization implementation tasks, and a semi-structured interview.

\subsection{Participants}
We recruited 16 participants (9 females, 7 males; P1--P16) through social media and word of mouth. All participants had prior experience with vibe coding tools and were capable of understanding and designing visualizations.
Participants were divided into two groups (8 each) based on their D3.js experience. The novice group had little to no D3 experience (at most one to two coursework charts), while the expert group had completed at least one non-trivial D3 project.
Participants came from diverse backgrounds, including data visualization, human--computer interaction, art \& design, microelectronics, and civil engineering.
Six participants were PhD students or held PhD degrees, and ten were master's students or held master's degrees.
% We recruited 16 participants (9 females, 7 males, denoted as P1 to P16) through social media and word of mouth under two preconditions: all participants had prior experience with vibe coding tools and were able to understand and design visualizations.
% On top of these preconditions, we split participants into two groups of 8 based on their D3.js implementation experience.
% The D3 novice group had no prior D3 coding experience or had only built one to two coursework charts.
% The D3 expert group had completed at least one D3 project with non-trivial customization.
% Participants came from diverse backgrounds, including data visualization, human-computer interaction, art \& design, microelectronics, and civil engineering.
% Among them, 6 held or were pursuing PhD, 10 master's degrees.

\subsection{Implementation Tasks and Materials}
\label{sec:tasks}
This study focuses on the visualization implementation phase. We assume participants begin with a well-formed design in mind and use a vibe coding tool to implement it in D3. To ensure comparability across expertise levels, we provided fixed target visualizations as proxies for user-intended designs.

\begin{figure}[t]
  \centering
  \includegraphics[width=\columnwidth]{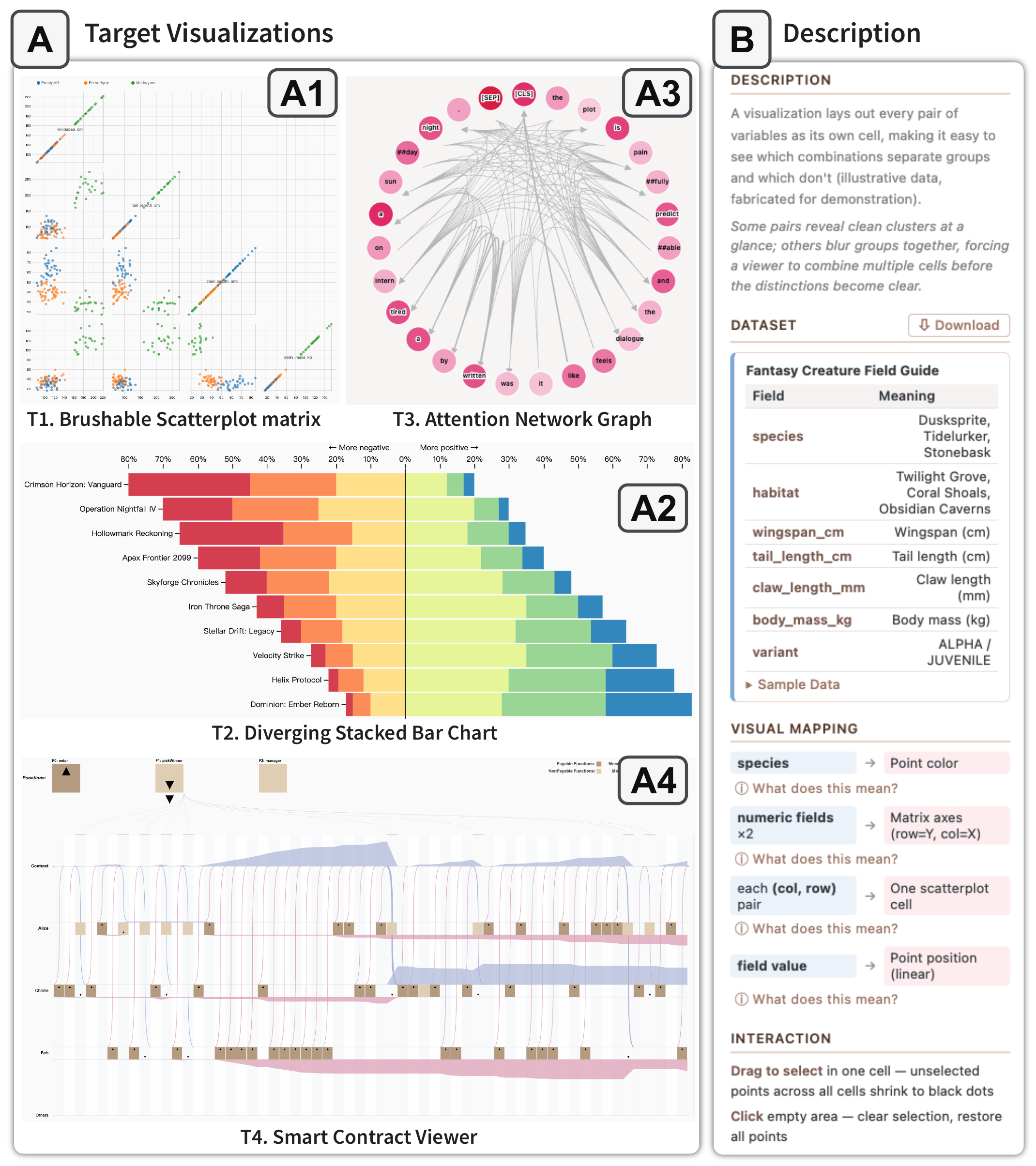}
  \caption{Materials for the user study tasks. We provide four target visualizations (A): two basic charts (A1, A2) and two customized visualizations (A3, A4). For each, we include a description (B) detailing the overview, dataset, visual mappings, and interactions.}
  \label{fig:task}
\end{figure}

We designed four target visualizations along two dimensions: complexity (basic vs. customized) and challenge type (visual mapping vs. interaction) (\autoref{fig:task}A). The two basic charts were selected from the D3 Observable Gallery~\cite{bostock_d3_gallery}: a diverging stacked bar chart (data transformation and multi-channel encoding) and a brushable scatterplot matrix (cross-view interaction). The two customized visualizations were adapted from prior systems: a smart contract viewer~\cite{wen2025prettismart} (composite glyph design) and an attention network graph~\cite{wang2021dodrio} (custom interaction on a radial layout). For customized cases, we simplified datasets and focused on a single primary view to reduce cognitive load.
We developed a presentation tool to display each target visualization with an introduction panel (\autoref{fig:task}B), including a chart description, a dataset description, visual mappings, and interactions. This helped participants quickly understand the design and focus on implementation. To avoid direct copying, screenshots and text copying were disabled, and participants were given datasets different from the examples.
To balance task coverage and workload, each participant completed one basic and one customized visualization. This ensured all four visualizations were covered while keeping the session manageable.

\subsection{Procedure}
Each participant attended a one-to-one session, either in person or remotely via Zoom, lasting 90--120 minutes. Before the session, we introduced the study procedure and obtained consent for screen and audio recording.
Each session comprised three phases: briefing, implementation tasks, and interview. In the briefing phase (5 min), the researcher explained the study purpose and introduced the visualization presentation tool. In the implementation phase (60 min), participants completed two tasks in order from basic to customized (\autoref{sec:tasks}), with optional breaks in between. Participants were informed that completing the visualization was not required, as we focused on their implementation process rather than final results. They used their own laptops and preferred vibe coding tools, interacting with the system using any input modality. After 30 minutes on each task, participants could choose to stop or continue.
The session concluded with a semi-structured interview (25 min) covering two parts: vibe coding practices and open discussion. The interview protocol targeted the end-to-end workflow, including goal formulation, prompting strategies, output evaluation, iterative refinement, and reflections on practices and challenges.

\subsection{Data Collection and Analysis}

We analyzed the collected data, which included participant conversation logs with vibe coding tools,
final code artifacts, transcribed interview responses, and screen recordings.
Each user prompt was coded against the reference model of
Card et al. \cite{card1999readings} as data transformation, visual mapping,
view transformation, or non-technical, with multiple labels allowed per prompt.
Initial labels were generated by an LLM and verified by the authors against the codebook.
Interview transcripts were then analyzed using inductive thematic
analysis to identify themes related to the two research questions.
In total, we coded 295 participant prompts and 16 interview transcripts;
all labels and themes were reviewed collaboratively by the research team.

\section{Results}

\begin{figure*}[t]
  \centering
  \includegraphics[width=\linewidth]{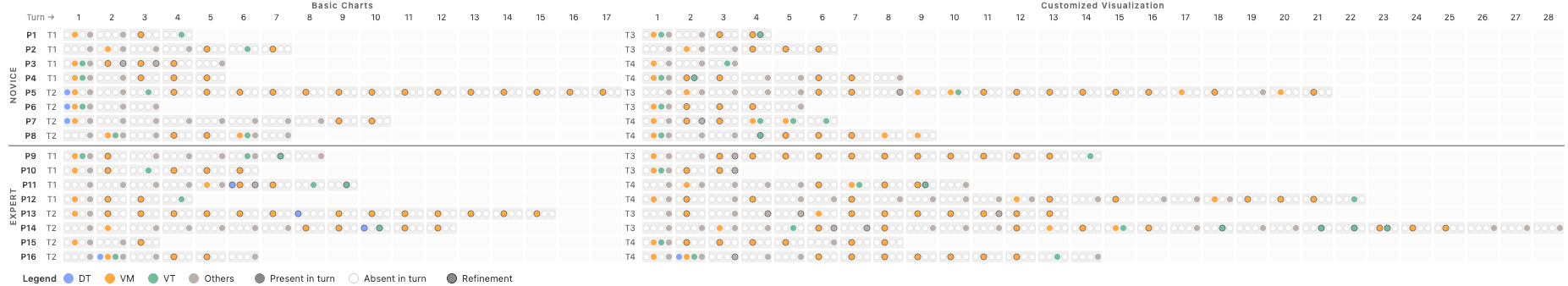}
  \caption{Participant Vibe Coding Practices for Visualization Implementation in the User Study. }
  \label{fig:bloch-sphere}
\end{figure*}

\subsection{Practice}

At the formulation stage, both expert and novice users wrote a long initial prompt that summarized the target visualization, typically covering data transformation and visual mapping.
As participants explained in the interviews, their goal was to obtain a runnable first version from a single prompt so that they could evaluate the AI's output visually rather than by reading the code.
Several participants (P12, P8, P4) also reported that the first prompt was harder to write for customized visualizations, since no common chart vocabulary applied to such designs.
\cref{fig:bloch-sphere} shows the per-turn category labels for each (participant, task) pair, with novices on the top and experts on the bottom.
A small number of participants instead opened with a Non-Technical only prompt, in which they asked the AI about the dataset structure or issued a short greeting or test message before moving to a substantive request.

Two formulation practices become visible along the turn axis of \cref{fig:bloch-sphere}.
In the \emph{static-first} practice, the participant scoped the initial prompt to data transformation and visual mapping only, drove the static chart to convergence, and only then introduced interaction, so view-transformation labels cluster at the right end of the row.
In the \emph{integrated} practice, the participant covered data transformation, visual mapping, and view transformation together in the first prompt and refined them one by one.
Across all turns, visual mapping dominates the conversation: visual mapping prompts account for the majority of turns and span the full length of nearly every session, whereas data transformation prompts are concentrated in the opening few turns and tend to be shorter than visual mapping prompts, partly reflecting our task design, in which the required data transformations were intentionally light.

\textbf{Prompt Patterns.}
Across both groups, participants prompted at the level of intent rather than implementation when specifying visual mapping. Novices reported that they could not give precise step-by-step instructions because they did not understand the underlying code logic, while experts reported that recent models had become strong enough that step-by-step instructions were no longer necessary in most cases. For data transformation requirements embedded in visual mapping prompts, participants typically wrote a single sentence stating the intended outcome and relied on the AI to identify the relevant fields and perform the transformation.

Specifying view transformation was the prompt type participants reported as most difficult to express in text. P4 explicitly noted that he had tried to describe the interaction word by word in the prompt, but the AI still failed to understand it. We observed two specification practices. The dominant one was behavioral description, in which participants stated the user-facing behavior of the interaction at a high level (e.g., \emph{``drag to select points within a chart, click empty area to restore''}). A second strategy was used by a small number of expert participants who were familiar with the underlying library: P11 embedded specific D3 API names such as \texttt{d3.circle} into the prompt, blending implementation-level vocabulary with natural-language requirements.

Text was the default modality for all participants, but nearly half of them switched to other modalities at moments when text was insufficient, and the trigger for each switch was specific to the kind of information that text could not capture. Spatial relationships and layout positions triggered hand-drawn sketches; for example, P2 reported wanting to draw by hand because spatial relationships were difficult to convey precisely in language. References to specific elements in an existing rendering triggered annotated screenshots (P5, P15). Customized visualizations without commonly known names sometimes triggered Figma prototypes (P12) or web searches for visually similar references. Besides, a few participants (P2, P16) pre-processed their prompts using a separate AI tool before sending the result to the vibe coding tool. For example, P16 drafted the prompt himself, then asked ChatGPT to clarify ambiguous wording and reorder the content into a more structured form.

\textbf{Evaluation.}
After receiving the AI's response, participants moved from writing prompts to checking whether the result matched their intended visualization.
In principle, this check could be carried out in three ways: reading the AI's textual output, inspecting the generated code, or running the code and examining the rendered output in the browser. 
Participants rarely reviewed the generated source code. 
Only three participants read the model's textual output, but most ran the code and judged the result through visual inspection and direct interaction with the rendered visualization. 
In the interviews, several participants (P2, P4, P12, P16) noted that they inspected the code only when the AI repeatedly produced results that did not meet their requirements, or when its modifications spanned multiple files.
Beyond these manual review practices, one expert participant attempted to fully automate the evaluation process by having the AI operate the browser and inspect the rendered output through screenshot feedback, covering both static charts and interactive behaviors.
% The limitations of this approach are discussed in Section~\ref{sec:challenges}.

Beyond cases where the generated code failed to meet the prompt's requirements, participants reported two recurring issues in the AI's output, which surfaced both when a single prompt carried many requirements and when conversations extended over multiple turns.
The first occurred during iteration: in satisfying the latest prompt, the AI sometimes violated constraints established by earlier prompts, so locally correct changes moved the visualization further from the intended design.
The second occurred when a single prompt contained multiple requirements: the AI often satisfied only a subset of them but reported in its summary that all had been completed.
These issues shaped how participants approached the iteration phase.

\textbf{Iteration.}
Novice and expert participants differed in their iteration practices.
Novices rarely edited the code themselves, citing their lack of D3 knowledge.
They iterated on the prompt instead and relied on the AI to eventually produce a workable result. 
They turned to manual editing only in extreme cases.
For example, P5, who had prior HTML experience, said that they would manually modify basic settings in the code when the AI repeatedly failed to produce the correct output.
In contrast, experts edited the code themselves for two main reasons.
First, when the AI became stuck in a loop of failed revisions, experts inspected and fixed the code directly to break out of the loop.
Second, for changes that were simple and well understood, such as adjusting fonts, axis endpoints, or color codes, experts edited the code directly because the AI round trip was slower than a manual edit.
As one expert put it, ``AI takes thirty seconds to change one line; I can do it in ten.''

Beyond this contrast in strategy, we also observed several techniques that participants adopted during refinement to communicate their intended changes more precisely.
Some participants provided feedback in the form of a screenshot of the incorrect visualization, often with annotations marking the specific elements to be changed, rather than describing the change in text.
Others used browser developer tools to extract the DOM id of a faulty element and passed this id to the AI, anchoring its attention on the corresponding code segment.

\subsection{General Questions}
This section reports feedback from general questions, including challenges, differences from other vibe coding scenarios, and sense-making in visualization implementation.
% In the interviews, we covered three open-ended questions that participants could answer based on both their experience in this study and their prior experience with vibe coding for visualization projects.

\textbf{Challenges.}
Beyond the recurring issues with the AI's conversational output already discussed in the Evaluation paragraph, our interviews surfaced two additional classes of challenges.

Although vibe coding substantially accelerates production compared with traditional line-by-line development, communicating intent to the AI remains a major bottleneck.
First, participants often struggle to express their intent clearly.
Natural language remains the most natural communication channel, and most participants only turn to other modalities when natural language alone proves insufficient; however, complex visual--spatial relationships are difficult to express precisely in natural language, since the medium itself imposes a limit.
This is especially the case for uncommon, customized visualizations, where participants do not even know how to articulate their requirements in natural language.
P13 noted that, for complex visualizations, communicating the intended design clearly enough that the vibe coding tool understands is harder than implementing it by hand.
Second, the AI lacked sufficient prior knowledge, while participants cannot mention every detail in their prompts; as noted in the Evaluation paragraph, common-sense application errors should be handled by the AI itself.
Third, the modalities supported by vibe coding tools are limited, which makes it difficult to communicate dynamic effects such as animation or interaction.
Some participants mentioned that being able to upload a short video or interactive demo would help.
In our study, participants could only convey static information through text, sketches, or prototype mockups, even though visualization interactions often involve transitional animation.
Although dynamic effects can be approximated by capturing multiple frames, this representation discards temporal information and is time-consuming to produce.

The second class concerns the efficiency of the vibe coding tool itself.
Based on the practices of expert users, vibe coding tools are slower than manual editing for small adjustments, yet visualization design typically requires many such fine adjustments to achieve a polished result.
In these cases, experts switched to direct code editing, while novices, lacking the ability to review code on their own, had to keep iterating on prompts, which reduced their efficiency.
Another efficiency issue is the long waiting time during AI generation.
Some participants expressed concerns about the time consumed while waiting for AI's outputs.
In our interviews, participants reported that during this waiting time they would either work on other parts of the same task or switch context to unrelated activities; only a few mentioned monitoring intermediate textual output to track whether the AI was working as intended.

\textbf{Differences from other vibe coding scenarios.}
We asked participants who had prior vibe coding experience in other domains, primarily software development and algorithm development, whether vibe coding for visualization is distinctive.
Their reflections drew on both their study experience and their prior development experience.

In terms of evaluation, participants noted that the success criteria of visualization are unique.
In conventional software development, AI output can be judged by binary signals such as ``the function runs'', ``the tests pass'', or ``the errors disappear''; in visualization, no such binary criterion exists, and the final judgment falls on whether the result ``looks right'' or ``looks good'', which currently can only be made by humans.
P4 and P9 emphasized that, in addition to logical correctness, visualization is also an aesthetic problem.
P5 added that, compared with software development, visualization lacks a clear-cut right-or-wrong judgment because visualization development is an iterative process and there is always a potentially better design.
P9 (expert) further noted, from a model-capability perspective, that ``the model itself currently has a hard time judging how aesthetic a given visualization actually is.''
These observations have two implications for the vibe coding workflow: (i) the AI cannot use ``tests pass'' as a convergence signal, so evaluation has to remain in the hands of the user, and (ii) users must visually inspect the chart at every iteration in order to decide whether to proceed, rather than relying on build status.

\textbf{Sense-making.}
All participants agreed that the visualization designs used in the study were reasonable and that they could extract meaningful insights from them.
However, some participants noted that interpreting the visualizations requires a certain level of literacy.
The customized visualizations used in the study had a non-trivial reading barrier, and participants suggested that, as designers, they would tend to add more onboarding content to lower the cognitive burden on viewers.

% \section{Discussion}

% \subsection{Significance}

\subsection{Implications for Visualization Implementation Tools}

Based on these challenges, we distill some design implications that follow directly from our participants' practices and the difficulties they encountered.

\textbf{Element-anchored editing.}
Vibe coding tools for visualization should treat individual visual elements as addressable units, so that a user can anchor a request on a clicked mark, axis label, or glyph component, rather than having to describe the element in text.
When the user clicks an element, the tool exposes two paths: for users who prefer to remain at the prompt level, the corresponding code segment and DOM identifier are forwarded to the AI as part of the next prompt; for users willing to edit code, the tool jumps directly to the relevant source line.
This design responds to two findings from our study.
On the communication side, visualization lacks the standardized vocabulary that programming has for naming code constructs, so users frequently struggle to specify which visual element they wish to change.
On the efficiency side, expert participants already approximate this workflow by hand.
P11 reported that ``once you pin it down to a specific DOM element and tell it exactly how to change it, the result is always better than a vague natural language description''.

Making visual elements directly addressable removes the missing vocabulary barrier for novices and gives experts a one-click path from the rendered chart to the source.

\section{Limitations and Conclusion}

To identify common patterns, we fixed the target visualizations and focused on the implementation phase, assuming the design and goals were predefined. This simplifies comparison but omits earlier design exploration and the co-evolution of design and implementation that often occurs in real-world vibe coding. In addition, we used clean, ready-to-use datasets, which did not reflect the complexity of real data preparation and transformation. Future work can study end-to-end workflows with open-ended tasks and more realistic datasets to better capture how users integrate vibe coding across the full visualization process.

In conclusion, we present an empirical study of visualization implementation through vibe coding, examining how users construct visualizations and iterate through prompts. Our findings reveal distinct practice patterns across data transformation, visual mapping, and view transformation, as well as key challenges in aligning intent, managing multi-step processes, and refining outputs. These insights inform the design of future tools that better support user-centered human--AI collaboration in visualization.


%\bibliographystyle{abbrv}
\bibliographystyle{abbrv-doi}

\bibliography{template}

\end{CJK}
\end{document}